\documentclass[twocolumn,showpacs,prl,floatfix]{revtex4}
\usepackage{graphicx}

\begin{document}

\title{Renormalization of the vacuum angle for a particle on a ring}

\author{S. M. Apenko }

\affiliation{ Theory Department, Lebedev Physics Institute,
Moscow, 117924, Russia}
 \altaffiliation[Also at ]{ ITEP, Moscow, 117924, Russia}
 \email{apenko@lpi.ru}

\date{\today}

\begin{abstract}
We analyze the vacuum (topological) angle $\theta$ renormalization
for the quantum mechanical model of a particle moving around a
ring, where $\theta$ is the magnetic flux through the ring. We
construct a renormalization group (RG) transformation for this
model and derive exact RG equations which lead to the flow diagram
similar to that of the Quantum Hall effect. Renormalization of
$\theta$ is seen to follow from the loss of information about the
initial topological charge in the course of the RG procedure.
\end{abstract}

 \maketitle

Topologically nontrivial field theories are of considerable
interest both in high energy and condensed matter physics
\cite{Pol}. Different non-linear sigma models with topological
terms describe e.g. the Quantum Hall effect (QHE)\cite{H},
antiferromagnetic spin chains \cite{S}, tunneling effects in
metallic nanostructures \cite{Z}(see also recent work \cite{B}).
Common feature of these models is the renormalization of the
vacuum angle $\theta$ \cite{KM,PB}, which is the coefficient in
front of the topological charge in the action. While generally
accepted by now, this $\theta$ renormalization still may seem
somewhat obscure. For this reason we study here how this
renormalization occurs in a simple quantum mechanical model (a
particle on a ring). After constructing a certain kind of the
renormalization group (RG) transformation it appears possible to
obtain analytically almost the same RG flow diagram as in the QHE
both for weak and strong coupling.

Consider a particle of mass $m$ moving around a ring of unit
radius threaded by a magnetic flux $\theta$ (in units
$c=\hbar=e=1$). The corresponding (euclidian) action at finite
temperature may be written in terms of a planar unit vector $\bf
{n}(\tau)$ (${\bf n}^2=1$) which depends on a one-dimensional
coordinate (imaginary time)
\begin{equation}\label{A}
  S[{\bf n}]=\frac{m}{2}\int_{0}^{\beta}\dot{\bf{n}}^2(\tau)d\tau-
  i\frac{\theta}{2\pi}\int_{0}^{\beta}\epsilon_{ab}n_{a}(\tau)\dot{n}_{b}(\tau)d\tau,
\end{equation}
where $\epsilon_{ab}$ is the two dimensional antisymmetric tensor
and $\beta$ is the inverse temperature.  Since ${\bf n}(0)={\bf
n}(\beta)$ the model is actually defined on a circle. The last
term in (\ref{A}) has the form $i\theta Q$ where $Q$ is the
topological charge which distinguishes inequivalent mappings
$S^1\rightarrow S^1$ and takes integer values (equal to a number
of  rotations  the particle make in time $\beta$), making the
theory periodic in $\theta$. This $(0+1)$-dimensional field theory
may seem trivial, since  in terms of the polar angle the action
(\ref{A}) is quadratic, but this is not so due to topological
effects (like in compact electrodynamics \cite{Pol}).

This model may be used also to describe a single electron box
(SEB)\cite{SEB}, which is essentially a metallic island coupled to
the outside circuit by a tunnel junction of capacitance $C$ and
resistance $R$. If $R\rightarrow\infty$ the action for the SEB
reduces to Eq. (\ref{A}) where the first term accounts for the
charging energy and $1/m=e^2/C$ while $\theta$ is an external
charge. Certainly, the nonlocal model for finite $R$
\cite{Z,B}(which also corresponds to the particle on the ring with
Ohmic friction \cite{G}) and at $C\rightarrow 0$ has more in
common with two-dimensional sigma models (e.g. dimensionless
coupling and asymptotic freedom \cite{Ko}, instantons of all sizes
\cite{Kor}), but the model of Eq. (\ref{A}) is much simpler and
seems to be most suitable for the analytical study of the vacuum
angle renormalization.

Since for the particle on the ring $\theta$  is an external
magnetic flux, one would normally expect its renormalization due
to the screening of the flux by the magnetic field, produced by
the rotating particle, but this mechanism does not work here,
because such back reaction is not included in Eq. (\ref{A}). We
shall see below, that quite a different ('informational')
mechanism is relevant here, related to the way  the topological
charge changes under the RG transformation.

Now we shall construct the  RG transformation for the action
(\ref{A}). For this purpose let us introduce a lattice dividing
the whole $\tau$ axis into intervals of the length $a$ and then
fix the values of the field $\bf{n}(\tau)$ at the sites $a_{i}$ of
the lattice. This results in a discrete configuration $[{\bf
n}_{i}]$ which will later play the role of a slowly varying
background field.

Next we  evaluate the 'probability' $P[{\bf n}_{i}]$ of this
configuration integrating out all remaining degrees of freedom,
i.e.
\begin{equation}\label{pro}
  P[{\bf n}_{i}]\sim \prod_{i}\int_{{\bf n}(a_{i})={\bf n}_{i}}
  ^{{\bf n}(a_{i+1})={\bf n}_{i+1}}D{\bf n}(\tau)
  \delta({\bf n}^{2}(\tau)-1)\exp(-S[{\bf n}])
\end{equation}
with the action $S$ from Eq. (\ref{A}). Obviously, this is just
the product of the Green functions $G({\bf n}_{i},{\bf
n}_{i+1};a)$ at the inverse temperature $a$, which are known
exactly for the simple model in question. If we introduce polar
angle $\phi$ instead of the planar vector then
\begin{eqnarray*}
  G(\phi_{i},\phi_{i+1};a)=\left(\frac{m}{2\pi a}\right)^{\frac{1}{2}}\sum_{q}
  \exp\left[-\frac{m}{2a}(\phi_{i+1}-\phi_{i}+\right.\\
  +\left.2\pi q)^2 +i\frac{\theta}{2\pi}(\phi_{i+1}-\phi_{i}+2\pi q)\right]
\end{eqnarray*}
where the sum is over different winding numbers $q=0\pm1\ldots$.
Then
\begin{equation}\label{p}
  P[\phi_{i}]\sim\prod_{i}G(\phi_{i},\phi_{i+1};a)\sim\exp(-S_{eff})
\end{equation}
with
\begin{eqnarray}\label{eff}
  S_{eff}=\frac{m}{2a}\sum_{i}(\phi_{i+1}-\phi_{i})^2-
  i\frac{\theta}{2\pi}\sum_{i}(\phi_{i+1}-\phi_{i})- \nonumber\\
  -\sum_{i}\ln\sum_{q}\exp\left[-\frac{2\pi^{2}m}{a}q^2
  +i\theta q -\frac{2\pi m}{a}q(\phi_{i+1}-\phi_{i})\right]
\end{eqnarray}
Up to now we did not specify the values of $\phi_{i}$ and these
were some arbitrary numbers. Now let us assume that adjacent
$\phi_{i}$  are very close to each other, i.e. they represent some
smooth  continuous field $\phi(\tau)$. Then we may write
$\phi_{i+1}-\phi_{i}\simeq \dot{\phi}a$ with
$\dot{\phi}\rightarrow 0$ and expand the effective action
(\ref{eff}) in terms of the derivative. The result of
straightforward calculations is given by the same formula Eq.
(\ref{A})
\begin{equation}\label{ef}
  S_{eff}[\phi]=\frac{m'(a)}{2}\int_{0}^{\beta}\dot{\phi}^2(\tau)d\tau-
  i\frac{\theta'(a)}{2\pi}\int_{0}^{\beta}\dot{\phi}(\tau)d\tau+\ldots
\end{equation}
but with renormalized coupling constants, which now depend on $a$
\begin{eqnarray}\label{ren}
  m'(a)=m-4\pi^2\frac{m^2}{a}\frac{\partial^2 f(\theta,a)}{\partial\theta^2},\nonumber\\
  \theta'(a)=\theta-4\pi^2\frac{m}{a}\frac{\partial f(\theta,a)}{\partial\theta},
\end{eqnarray}
where
\begin{eqnarray}\label{F}
  f(\theta,a)=-\ln Z(\theta,a),\qquad\qquad \nonumber\\
  Z(\theta,a)=\left(\frac{m}{2\pi a}\right)^{\frac{1}{2}}
  \sum_{q}\exp\left[-\frac{2\pi^2m}{a}q^2
  +i\theta  q\right].
\end{eqnarray}
Here $Z(\theta,a)$ is the partition function  for the particle on
the ring (at inverse temperature $a$) represented as a sum over
winding numbers.

If we introduce a dimensionless coupling constant
\begin{equation}\label{g}
  g=\frac{a}{m}
\end{equation}
we can easily see from Eq. (\ref{ren}) and Eq. (\ref{F}) that both
$\theta'(a)$ and $g'(a)=a/m'(a)$ depend on the scale $a$ only
through $g$. This means that derivatives of $g'$ and $\theta'$
with respect to $\ln a$ can be expressed solely through the
running couplings (after inverting Eqs. (\ref{ren})) to obtain RG
equations in more familiar form
\begin{equation}\label{rg}
  \frac{dg'}{d\ln a}=\beta_1 (g',\theta'),\qquad
  \frac{d\theta'}{d\ln a}=\beta_2 (g',\theta')
\end{equation}
However, we do not actually need $\beta$-functions here, since we
already have the solutions of these RG equations (in terms of
initial values $m$ and $\theta$) given by Eq. (\ref{ren}).

Let us now comment on the physical meaning of the transformation
constructed. While this is close in spirit to the real space RG
approach here we do not eliminate fast variables step by step. It
is difficult to find a suitable decomposition of fields into fast
and slow components in theories with constraints. Usually, in
sigma models with ${\bf n}^2=1$ one adopts the RG scheme due to
Polyakov \cite{Pol}, but this describes only small fast vibrations
of a slowly rotating unit vector, i.e. fast variables are
topologically trivial. But as we know (and will see later) fast
rotations are also to be taken into account, if one wishes to
obtain $\theta$ renormalization. These fast rotations are usually
represented by instantons of small size \cite{KM,PB}, but this
makes sense only in the weak coupling region.

What is done here is in fact the same as the preliminary step for
the Wilson's RG, when he introduces a smooth average order
parameter $M({\bf x})$ \cite{W}, with Fourier components lower
than some ultraviolet cutoff and defines an effective action for
$M({\bf x})$ performing a statistical averaging holding $M({\bf
x})$ fixed for all ${\bf x}$. The only (unimportant) difference is
that here we take a discrete set $[{\bf n}_{i}]$ as an order
parameter, and only later take the continuous limit.
 It is not so obvious, however, how to use the
effective action (\ref{ef}), since as we shall see later, at
$\theta\neq\pi$ the effective mass $m'(a)$ goes to zero as
$a\rightarrow\infty$ and fluctuations become large.  Therefore, if
we have in mind further functional integration over the slow field
$\phi(\tau)$ we should take account of the higher powers of
$\dot{\phi}$ (and higher derivatives).

There exists however a different interpretation which seems to be
more instructive. One can view  the procedure described as a kind
of a continuous measurement. In fact, $P[{\bf n}_{i}]$ is an
amplitude for a process when successive measurements of the unit
vector directions at times $a_{i}$ give the values ${\bf n}_{i}$.
One may think that we look at our system at discrete time moments
and make a series of snapshots which are then combined into a
'movie'. This cinematic sequence (similar to the 'coarse-grained
history' of Gell-Mann and Hartle \cite{GH}) looks like a smooth
trajectory and the corresponding amplitude is determined by the
effective action (\ref{ef}). This interpretation provides us with
additional physical meaning of parameters $m'(a)$ and $\theta'(a)$
--- they characterize the slow motion of a particle being
continuously measured by some external observer (or environment)
with time resolution $a$.

Between the measurements the particle moves freely and can even
perform many revolutions around the ring which we are unaware of.
To state this more rigorously, we are always mistaken when we
ascribe a certain topological charge $Q'$ to a coarse grained
field. The true topological charge $Q$ should include also a
number of fast rotations (topologically nontrivial short
wavelength fluctuations, e.g. instantons and anti-instantons)
which we are unable to resolve. Then, to reproduce the true phase
factor of a given field configuration $\exp(i\theta Q)$ we have to
change the value of the vacuum angle $\theta$ and to make it scale
dependent. This is essentially what is done when passing from Eq.
(\ref{A}) to Eq. (\ref{ef}). Surely, many  initial field
configurations with different topological charges result in the
the same coarse grained field after averaging over fast variables,
i.e. the RG transformation mixes different topological sectors of
the theory. This lost information about the original values of $Q$
is now encoded in the renormalized angle $\theta'(a)$, hence this
renormalization may be said to be of 'informational' origin.

Let us study now Eqs. (\ref{ren}) in more detail. First of all
consider the case of small $a$. In this limit the terms with large
winding numbers in $Z(\theta,a)$ (\ref{F}) are strongly suppressed
and it is possible to retain only the terms with $q=0,\pm 1$. Then
the RG equations (\ref{ren}) for $g\ll 1$ can be written as
\begin{eqnarray}\label{rens}
  \frac{1}{g'}=\frac{1}{g}-\frac{8\pi^2}{g^2}{\rm e}^{-\frac{2\pi^2}{g}}\cos{\theta},\nonumber\\
  \theta'=\theta-\frac{8\pi^2}{g}{\rm e}^{-\frac{2\pi^2}{g}}\sin{\theta}.
\end{eqnarray}
It is possible also to rewrite this equations in the differential
form (\ref{rg}) with $\beta$ functions (here we omit the primes)
\begin{eqnarray}\label{b}
  \beta_1(g,\theta)=g-g^2D(g){\rm
  e}^{-\frac{2\pi^2}{g}}\cos{\theta},\nonumber\\
  \beta_2(g,\theta)=D(g){\rm e}^{-\frac{2\pi^2}{g}}\sin{\theta},
\end{eqnarray}
where $D(g)=8\pi^2(1-2\pi^2/g)/g^2$.

These formulas strongly resembles the instanton induced
renormalization  in the Yang-Mills theory \cite{KM} or in the QHE
\cite{H}. It should be stressed however that in our model
(\ref{A}) there are no instantons (as solutions of classical
equations) of small size. Nevertheless the physics is just the
same. Exponentially small contributions in (\ref{b}), coming from
the terms with winding numbers $q=\pm 1$, arise from the fast
instanton-like rotations of the vector ${\bf n}$ between the time
moments $a_i$ and $a_{i+1}$. Note that the same fluctuations are
responsible also for the mass renormalization  since without the
topological effects the model (\ref{A}) is trivial. The first term
in $\beta_1(g,\theta)$ which looks like a perturbative
contribution is in fact due to the explicit scale dependence of
the coupling constant $g$.

In the opposite limit of the strong coupling $g\gg 1$ we may use
the dual representation of the partition function $Z(\theta,a)$ as
a sum over the energy levels
\begin{equation}\label{Z}
  Z(\theta,a)=\sum_{n}\exp\left[-\frac{a}{2m}
  \left(n-\frac{\theta}{2\pi}\right)^2\right],
\end{equation}
where  $n=0,\pm 1\ldots$ is the angular momentum of the particle
measured in units of $\hbar$. Then, at $a\rightarrow\infty$ only
the ground state contributes to $Z(\theta,a)$, i.e.
\begin{equation}\label{z}
  Z(\theta,a)\simeq\exp(-aE_0(\theta)),
\end{equation}
where $E_0$ is the ground state energy for the particle on the
ring
\begin{eqnarray}\label{eo}
  E_0(\theta)&=&\frac{1}{2m}\left(\frac{\theta}{2\pi}\right)^2,\qquad
  0<\theta\leq\pi, \nonumber\\
E_0(\theta)&=&\frac{1}{2m}\left(1-\frac{\theta}{2\pi}\right)^2,\qquad
\pi<\theta<2\pi
\end{eqnarray}
Recall that there is level crossing at $\theta=\pi$ and the ground
state is degenerate at this point. Substituting these formulas
into Eq. (\ref{ren}) we immediately obtain
\begin{eqnarray}\label{ired}
  m'(a)&\rightarrow  & 0,\qquad  \theta\neq\pi, \nonumber\\
  \theta'(a)&\rightarrow &\left\{
\begin{array}{ll}
  0, & 0<\theta <\pi \\
  2\pi, &  \pi<\theta<2\pi
\end{array}\right.
\end{eqnarray}
Hence, in the long wavelength limit $\theta'(a)$ tends to a
stepwise function, just like the Hall conductivity in the QHE,
while $m'(a)$ vanishes like the diagonal conductivity. The
calculation at $\theta=\pi$ is slightly more involved (the
degeneracy of the ground level should be taken into account) and
results in a linear growth of the effective mass with $a$
according to $m'(a)\sim a/4$. Therefore, when expressed in terms
of $g$ the RG flow has a fixed point $\theta^*=\pi$, $g^*=4$.

To evaluate corrections to Eqs. (\ref{ired}) one has to take into
account also  the first exited level  and after some algebra we
obtain at $g\rightarrow\infty$
\begin{equation}\label{cor}
\frac{1}{g'}\simeq\exp(-g\Delta),\qquad \theta'\simeq
2\pi\exp(-g\Delta),
\end{equation}
at $0<\theta<\pi$ (for $\theta$ not too close to zero) where
$\Delta =(1-\theta/\pi)/2$ is the dimensionless energy gap.
Therefore deviations from the ideal quantization at
$\theta\neq\pi$ are exponentially small at large scales as it
should be in a system with the finite correlation length $\xi\sim
m\Delta^{-1}$. Moreover, at $g\gg 1$ running couplings are related
by a simple linear relation
\begin{equation}\label{rel}
  \frac{1}{g'(a)}=\frac{1}{2\pi}\theta'(a).
\end{equation}

\begin{figure}[ht]
\includegraphics[width=3.375in]{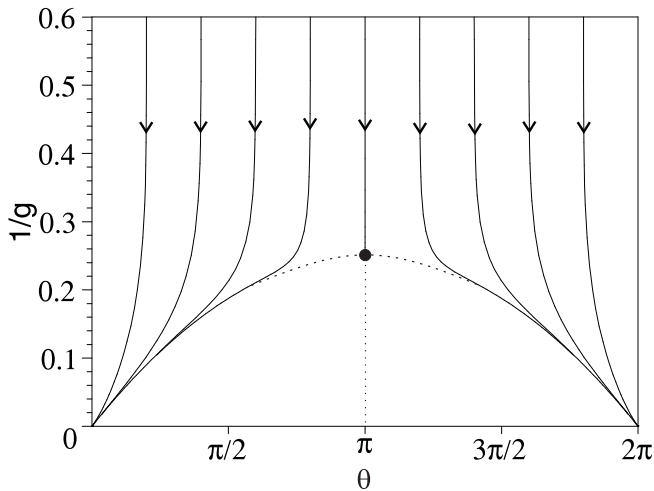}
\caption{RG flow diagram of Eqs. (\ref{ren}).} \label{fig}
\end{figure}

The whole RG flow diagram  is shown in Fig. \ref{fig} in the
$(1/g,\theta)$-plane for $0<\theta<2\pi$. The region below the
fixed point is not available in our model, since for a finite
initial $m$ RG trajectories in Fig. \ref{fig} always start at
infinity.

This is essentially the same diagram as proposed initially for the
integer Quantum Hall effect \cite{Kh,H}. Its main features are:
renormalization of $1/g$ to zero except for $\theta=\pi$ where the
saddle point exists at some finite $g^*$ and a set of infra-red
fixed points at quantized values of $\theta=0,\pm2\pi,\ldots$. Our
example shows that one can obtain such a non-trivial RG flow even
for a very simple system provided the RG transformation is
complicated enough. At a finite temperature $T$ the RG flow should
be stopped at $a=1/T$ since this is the size of the system
(\ref{A}) along the $\tau$ axis.

For the model considered the vanishing of the effective particle
mass (or $1/g$) as $a\rightarrow\infty$ is almost trivial. When
the scale $a$ becomes larger than the correlation length $\xi\sim
m \Delta^{-1}$,  adjacent ${\bf n}_i$ are no longer correlated
which results in the loss of the corresponding stiffness, i.e.
$m(a)\rightarrow 0$. From the point of view of one-dimensional
statistical mechanics this means that the RG flow leads us further
into the disordered phase (leading to the entanglement loss along
the RG trajectories \cite{E}). In the vicinity of $\theta=\pi$ the
correlation length diverges $\xi\sim 1/|\theta-\pi|$ and the fixed
point $g^*$ results from the degeneracy of the ground state at
$\theta=\pi$.

The physical meaning of the renormalized vacuum angle $\theta'(a)$
may be easily seen from Eq. (\ref{ren}). If we substitute
$Z(\theta,a)$ from Eq. (\ref{Z}) into Eq. (\ref{ren})  then
\begin{equation}\label{ch}
 \theta'(a)/2\pi=\langle n\rangle=\left\langle
 -i\frac{d}{d\phi}\right\rangle,
\end{equation}
where for the particle on the ring $n$ is the canonical angular
momentum and the brackets denote averaging over the ensemble with
the temperature $T=1/a$. For the SEB $\langle n\rangle$ is the
average number of excess electrons in the box and hence
quantization of $\theta'$ at $a\rightarrow\infty$ is just the zero
temperature Coulomb blockade \cite{AL} (for the SEB with $1/R\neq
0$ this was recently pointed out in \cite{B}). Thus the $\theta$
renormalization may be directly observed here (as in the QHE) as
the formation of the stepwise dependence of $\langle n\rangle$ on
the external charge $\theta$ ('Coulomb staircase') when the
temperature is lowered.

In summary, we have studied the $\theta$ angle renormalization for
the quantum mechanical particle moving around the ring, where
$\theta$ is the magnetic flux through the ring. The appropriate RG
transformation was constructed which resulted in the RG flow shown
in Fig.\ref{fig}. This flow diagram is similar to that of the
Quantum Hall effect and gives one more example of the $\theta$
renormalization.  We argue that this renormalization occurs due to
the information loss, because the RG transformation which
eliminates fast fluctuations from a given field configuration
changes its topological charge $Q$ (mixes different topological
sectors). Then, looking at the resulting coarse-grained field we
see different topological charge than the true one, so that to
recover the phase factor $\exp(i\theta Q)$ we have to change the
value of $\theta$ and to make it scale dependent. The absence of
the phase factor in the long wavelength limit (due to $\theta$
renormalization) may probably be viewed  as a complete screening
of the background topological charge by fast instanton-like
fluctuations.

I am grateful to V. Losyakov, A.Marshakov, A. Morozov, D.
Nurgaliev, I. Tipunin  and A.D. Zaikin for valuable discussions.
The work was supported  by the RFBR grant $\#$ 06-02-17459.

\end{document}